\documentclass{article}
\usepackage{ijcai24}
\usepackage{hyperref}
\usepackage{bm}
\usepackage{mathptmx}
\usepackage{amsmath}
\usepackage{graphicx} 
\usepackage{enumitem}
\usepackage[margin=1in]{geometry}
\usepackage{amsthm}
\usepackage{thmtools}
\usepackage{thm-restate}
\usepackage{amssymb}
\usepackage{mathtools}
\usepackage{subcaption}
\usepackage{mathtools}
\usepackage{lipsum}
\newtheorem{theorem}{Theorem}
\usepackage[ruled,noline,linesnumbered]{algorithm2e}
\newtheorem{lemma}{Lemma}
\newtheorem{proposition}{Proposition}

\DeclarePairedDelimiter\ceil{\lceil}{\rceil}
\DeclarePairedDelimiter\floor{\lfloor}{\rfloor}

\DeclareMathOperator*{\argmin}{argmin}  
\newcommand{\wdef}{W^{\rm l}}
\newcommand{\watt}{W^{\rm f}}
\newcommand{\minuseq}{\mathrel{{-}{=}}}
\newcommand{\pluseq}{\mathrel{{+}{=}}}

\title{Adversarial Knapsack for Sequential Competitive Resource Allocation}
\author{Omkar Thakoor, Rajgopal Kannan, Victor Prasanna}
\date{}

\begin{document}

\maketitle

\begin{abstract}
This work addresses competitive resource allocation in a sequential setting, where two players allocate resources across objects or locations of shared interest. Departing from the simultaneous Colonel Blotto game, our framework introduces a sequential decision-making dynamic, where players act with partial or complete knowledge of previous moves. Unlike traditional approaches that rely on complex mixed strategies, we focus on deterministic pure strategies, streamlining computation while preserving strategic depth. Additionally, we extend the payoff structure to accommodate fractional allocations and payoffs, moving beyond the binary, all-or-nothing paradigm to allow more granular outcomes.

We model this problem as an adversarial knapsack game, formulating it as a bilevel optimization problem that integrates the leader’s objective with the follower’s best-response. This knapsack-based approach is novel in the context of competitive resource allocation, with prior work only partially leveraging it for follower analysis. Our contributions include: (1) proposing an adversarial knapsack formulation for the sequential resource allocation problem, (2) developing efficient heuristics for fractional allocation scenarios, and (3) analyzing the 0-1 knapsack case, providing a computational hardness result alongside a heuristic solution.

\end{abstract}

\section{Introduction}

Strategic competition where two players allocate resources across objects or locations of shared interest, is a commonly occurring scenario. The Colonel Blotto game is a popular framework for such scenarios. The original model introduced in \cite{borel1953theory} has two players assigning resources (troops) across multiple battlefields, aiming to maximize their number of victories. The winner of a battlefield is determined by having more troops than the opponent, and the final payoff is the number of battlefields won. Initially developed for military strategy, the model has been applied to various competitive contexts, such as sports, advertising, and politics.

In its classical form, the solutions are given by Nash equilibria where both players \emph{best respond} to the opponent strategies. In two-player zero-sum games, Nash equilibria are shown equivalent to MaxMin strategies ensuring optimal worst-case outcomes for rational players. However, finding these strategies in Colonel Blotto games is computationally complex, as their solution space can be encoded in simpler two-player games but requires significantly more computation due to the nature of the payoff structure and strategic interactions.

Our work distinguishes itself from traditional Colonel Blotto games in three key aspects. First, while the Colonel Blotto framework operates in a simultaneous setting where players allocate resources without observing the opponent's actions, our approach considers a sequential setting, introducing a dynamic where decisions are made with partial or complete knowledge of previous moves. Second, instead of relying on complex mixed strategies that involve probabilistic allocations, our model emphasizes the use of deterministic pure strategies, simplifying strategy computation while maintaining effectiveness in decision-making. Third, we extend the payoff structure to allow for fractional allocation and fractional payoffs, moving away from the traditional all-or-nothing paradigm where victory in a battlefield is entirely binary, thereby enabling a more nuanced evaluation of outcomes.
Our approach is to model this as a knapsack game, and in particular a bilevel optimization problem for the adversarial knapsack formulation. While a natural fit for this scenario, the knapsack perspective has been seldom used or referred at all by previous works, with the only exception \cite{behnezhad2018battlefields} limiting its use for the follower best-response, not extending it to the leader objective as we do.

Our contributions are as follows: We propose an adversarial knapsack formulation for the sequential competitive resource allocation problem. We analyze the fractional allocation setting providing efficient heuristics. Finally, we also analyze the conventional all-or-nothing scenario under 0-1 knapsack formulation providing computational hardness result along with a heuristic solution.

\section{Related work}
The classical Colonel Blotto and vast majority of its variants consider a zero-sum game, but its complexity arises from the exponential growth of pure strategies with the number of troops and battlefields, making the computation of optimal strategies challenging. Since its inception,
\cite{tukey1949problem,blackett1954some,blackett1958pure,bellman1969colonel,shubik1981systems,roberson2006colonel,kvasov2007contests,hart2008discrete,golman2009general,kovenock2012coalitional,weinstein2012two}, 
various approaches have addressed specific cases of the problem. Many studies relax the integer constraint, exploring a continuous version where troops are divisible. For instance, \cite{borel1991application}
provided the first solution for three battlefields, while \cite{RM-408}
extended this to any number of battlefields, assuming equal troop numbers for both players. \cite{roberson2006colonel}
examined optimal strategies in symmetric games where all battlefields have equal weight. \cite{hart2008discrete}
tackled the discrete, symmetric version for specific cases. Recently, \cite{ahmadinejad2019duels}
made significant progress by devising optimal strategies using exponential-sized linear programs and employing the Ellipsoid method to achieve polynomial-time solutions, gaining widespread attention. However, while of great theoretical implication, their algorithm
is not of much practically value as its computational complexity is $O(B^{12}N^4)$ where B is the players' resource budget.

In recent years, there has been significant exploration of bilevel variations of the knapsack problem. For instance, \cite{dempe2000bilevel} examined a model where the leader determines the knapsack's weight capacity, while the follower selects which items to include within the given constraint. A different approach is proposed by \cite{mansi2012exact}, who studied a bilevel knapsack problem in which the item set is divided between the leader and the follower, with each player managing their respective subsets. Another variation, introduced by \cite{denegre2011interdiction}, involves both players having their own knapsack, with the follower restricted to choosing only from the items that remain unpacked by the leader. \cite{caprara2014study} show the three aforementioned variants to be $\Sigma_2^P$-complete.
A key distinction in our problem that poses a challenge is that the inner level problem has bilinear terms consisting of products of outer and inner variables.

\section{Adversarial Knapsack}

Recall that an instance of the knapsack problem consists of a set of items with given weights and profits together with a knapsack with a given weight capacity. The objective is to select a subset of the items with maximum total profit, subject to the constraint that the overall selected item weight must fit into the knapsack.

We consider a game between two players \textemdash{} leader $\mathrm{l}$ and follower $\mathrm{f}$ \textemdash{} allocating resources to a set of $n$ items.
Each item $i$ has value $v_i$, and we denote the vector of all values as $\bm{v}$. Leader sets a weight $w_i$ for each $i$ which represents how much resources the follower would need, to \emph{fully win} item $i$.
For the player's strategies and payoffs, we consider two different settings. 
In the first, we consider a 0-1 knapsack to capture an all-or-nothing scenario, i.e., if the follower decides to win an item, they must commit resources worth $w_i$. The second setting is that of fractional knapsack, where allocating fewer resources than the set price will yield a proportionally fractional utility. In practice, this can be perceived as the expected utility, corresponding to a likelihood of winning the contest being the fraction of required allocation.
Since the leader resources are limited, we represent it with a budget parameter $\wdef$ and require $\sum_i w_i \le \wdef$. The follower has a knapsack capacity $\watt$ representing its available resources.

Thus, our adversarial knapsack game is described by the tuple $\langle n, \bm{v}, \wdef, \watt \rangle$. Follower aims to compute max-valued knapsack given leader's weight assignment, and the leader wants to set weights $\bm{w}$ to minimize the follower's payoff.
Our analysis shows how the problem complexity varies when the leader can only set integer weights, i.e., from a discrete space versus when they are allowed to be any real values, i.e., from a continuous space.

\section{Fractional Formulation}

In the zero-sum game, we do not need to distinguish between strong vs weak Stackelberg Equilibrium, leading to the following bi-level Optimization Problem (OP) for the leader:
\begin{align*} \label{DefWSENoUnc}
\min_{\sum w_i \le \wdef} \ \ \max_{x} \ & \sum_i v_ix_i \\
\text{s.t.} \ & \sum_i w_ix_i \le \watt \\
& x_i \in [0,1]
\end{align*}

The inner level is for the follower's best response which is to solve the knapsack problem, and the outer level is for the leader adversarially setting optimal weights.

For the follower, Dantzig's algorithm is a well-known greedy method to achieve the maximum knapsack value. It simply picks items in the order of their \textit{bang-for-the-buck} $\frac{v_i}{w_i}$ (\textit{BB} hereafter), from highest to the lowest, until the knapsack is full.

When the leader can set non-integer weights, we have a solution given by the following proposition:

\begin{proposition}
    The leader's optimal solution solution is to set each $w_i$ proportional to its $v_i$, i.e., $w_i = \frac{v_i}{\sum_i v_i} \wdef \ \forall i$.
\end{proposition}

\begin{proof}
    We note that the follower is always able to exhaust their budget in the continuous knapsack setting, so given any weights $\bm{w}$, their payoff is $\watt \cdot b$ where $b$ is the (weighted) average of the BB of the items they pick. But since they pick items in the order of highest BB first, it must be that $b \ge \tilde{b}$, where $\tilde{b}$ is the weighted average of BB of all the items. But since the total value and the total weight of all the items is fixed, so is $\tilde{b}$ ($= \frac{\sum_i v_i}{\wdef}$), irrespective of $\bm{w}$. So, the follower's payoff is minimized if the defender can achieve $b = \tilde{b}$, which is achieved by setting each item's BB to $\tilde{b}$, i.e., $\frac{v_i}{w_i} = \frac{\sum_i v_i}{\wdef}$, proving the result.
\end{proof}

We refer to this solution as the \emph{value-proportional} (VP) weights.
It follows that the VP weights might not be integers even when the item values and the budgets are. Hence, if $\bm{w}$ is restricted to all integers, the solution above does not apply. We investigate this setting next.

\subsection{Discrete resources}
Given the direct closed-form solution in the continuous case, the natural remedy to try for the discrete case is to obtain the VP weights and round them to the closest integer. It turns out that this does not retain optimality. 

\begin{figure}[t]
  \centering
  \includegraphics[width=.8\linewidth]{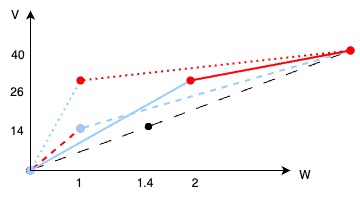}
  \caption{\small Example for discrete fractional setting showing follower payoff varies with follower budget $\watt$ for different weights $\bm{w}$. Item 1 ($v_1 = 1.4$) and item 2 ($v_2 = 2.6$) in red, blue resp. Weights (1,3), (2,2), (3,1) shown via dashed, solid and dotted lines resp. with continuous allocation baseline in black.}
  \label{fig:roundingCounter}
\end{figure}

Consider the example as follows. Suppose we have two items with values $14,26$, and say $\wdef=4$. The VP weights are $(1.4, 2.6)$ respectively, rounding which yields $(1, 3)$. For $\watt=2$, it yields $22.67$ for follower (greedily picking value 14 with weight 1 for item 1 followed by 1/3 of 26 (for item 2) with the remaining weight 1). This can be seen to be better than other solutions indeed. However, for $\watt=1$, the optimal is $(2,2)$ which restricts the follower payoff to 13 compared to the solution $(1, 3)$ which yields 14, thus, the rounded VP weights are not optimal. A key takeaway from the example as can be seen from Fig. \ref{fig:roundingCounter}, is that the optimal weights depend on $\watt$. Thus, any approach finding a good integer solution from a good fractional solution cannot be oblivious to $\watt$.

For our remaining results, we define a piecewise linear function $U_{\bm{w}}()$ for any $\bm{w}$, that gives the follower payoff in terms of follower allocation (which is what Dantzig's algorithm computes). E.g., fig. \ref{fig:roundingCounter} shows four different functions, with the VP weights yielding a linear function. When $\bm{w}$ are sorted in non-ascending BB of items, and $k$ is the largest index s.t. $w_1 + w_2 \ldots + w_k \le W$, we can write 

\[
U_{\bm{w}}(W) = 
\sum_{i=1}^k v_i + \frac{v_{k+1}}{w_{k+1}} \cdot (W - \sum_{i=1}^k w_i)
\]

It is also worth noting that we can construct examples where the optimal weights go beyond the floor and ceiling values of the corresponding VP weights, which suggests the problem to be even harder.
Suppose $\bm{w^{\rm in}}$ denotes the optimal solution with integer-restricted weights and $\bm{w^{\rm fr}}$ being the VP weights. Each $w^{\rm in}_j$ is either deflated or inflated relative to the corresponding $w^{\rm fr}_j$. We show that

\begin{lemma}\label{lem:flcl}
    $\bm{w^{\rm in}}$ satisfies either
    \begin{itemize}
        \item $w^{\rm in}_j \le w^{\rm fr}_j \ \Rightarrow \ w^{\rm in}_j = \floor*{w^{\rm fr}_j} \quad \forall j$ \qquad or, 
        \item $w^{\rm in}_j \ge w^{\rm fr}_j \ \Rightarrow \ w^{\rm in}_j = \ceil*{w^{\rm fr}_j} \quad \forall j$
    \end{itemize}
\end{lemma}

\begin{proof}[Proof sketch]
     Suppose we have weights $\bm{w}$ s.t. $\exists j, k$ with $w_j < \floor*{w^{\rm fr}_j}$ and $w_k > \ceil*{w^{\rm fr}_k}$. Then, we consider $\bm{w'}$ obtained by changing $\bm{w}$ s.t. $w'_j = w_j + 1$ and $w'_k = w_k -1$. Using the comparison of the BB values $\frac{v_j}{w_j} > \frac{v_j}{w'_j} > \frac{v_k}{w'_k} > \frac{v_k}{w_k}$ and the concavity of $U_{\bm{w}}()$ and $U_{\bm{w'}}()$, we are able to show that $U_{\bm{w}}() \ge U_{\bm{w'}}()$ (details omitted for brevity). Thus, we are able to change the weights while improving the solution until we can't find a pair $(j,k)$ as above, thus, the final solution $\bm{w^{\rm in}}$ then satisfies one of the two conditions in our result.
\end{proof}

Next, we analyze the worst-case difference in defender's objective due to the integer-restriction on weights.
\begin{proposition}
For the optimal solution $\bm{w}$ and VP weights $\bm{\tilde{w}}$, we have $U_{\bm{w}}(\watt) \le U_{\bm{\tilde{w}}}(\watt) \left(1 + \frac{n-1}{\watt} \right)$
\end{proposition}

\begin{proof}
    Recall that $U_{\bm{\tilde{w}}}$ is a linear function with a slope $\tilde{b} = \frac{\sum_i v_i}{\wdef}$. Next, the concave function $U_{\bm{w}}$ has the same value as $U_{\bm{\tilde{w}}}$ at $W=0$ and $W = \wdef$. Further, since $U_{\bm{w}}$ is piecewise linear, the function ($U_{\bm{w}} - U_{\bm{\tilde{w}}}$) is maximum at a breakpoint of $U_{\bm{w}}$ i.e., $W = \sum\limits_{i=1}^k w_i$ for some $k<n$. This simplifies to
    \[(U_{\bm{w}} - U_{\bm{\tilde{w}}})(\sum\limits_{i=1}^k w_i) = \tilde{b} \cdot \sum\limits_{i=1}^k (\tilde{w}_i-w_i) \le (n-1)\tilde{b}\]
    The bound is obtained from lemma \ref{lem:flcl}. Since this is a bound on the maximum, we get $(U_{\bm{w}} - U_{\bm{\tilde{w}}})(\watt) \le (n-1)\tilde{b}$ in particular. Rewriting $U_{\bm{\tilde{w}}}(\watt) = \tilde{b}\watt$ and rearranging terms gives us the result.
\end{proof}

\subsubsection*{Heuristic solutions}

Having analyzed the bounds and properties of the optimal solution, we now propose efficient heuristics which we evaluate numerically in section 5.

Our heuristics compute the optimal $\bm{w}$ by modifying the VP weights $U_{\bm{\tilde{w}}}$, constructing the functions $U_{\bm{w}}$ one piece at a time.
The first heuristic BBup, shown in Algorithm \ref{alg:bbup} works as follows. We iteratively consider items as a candidate first piece of $U_{\bm{w}}$ (Line 3) \textemdash{} for item $i$, we set $w_i = \floor*{\tilde{w}_i}$. The quality of solution $U_{\bm{\tilde{w}}}(\watt)$ after eventually setting $w_{-i}$, is (under-)estimated by considering it as one piece, i.e. with an average BB of the remaining items (hence the 'BB' in the heuristic name) resulting in a budget-proportional payoff for the follower (Line 4). The item with the best such estimate (Line 6) is picked as the first piece, and then removed from consideration (Line 7-8). The procedure is then repeated for the residual problem (Line 9). The procedure is halted when the assigned weights exceed $\watt$ and the remaining pieces are constructed similarly but in arbitrary order (details omitted for brevity).

While the previous approach constructs $U_{\bm{w}}$ \emph{going up}, we analogously consider BBdown that constructs the pieces \emph{going down}, and BB+ as the one that considers the better of the two. We discuss the advantages of doing this with our numerical results. It is easy to see that these have a runtime complexity of $O(n^2)$ due to $O(n)$ recursive calls each taking $O(n)$ time.

\begin{algorithm}[h]
\DontPrintSemicolon
\SetKwProg{Def}{}{:}{}
\Def{BBup $(n, \bm{v}, \wdef, \watt)$}{
$\bm{\tilde{w}} =$ VP weights\;
\For{$i = 1, 2, \ldots, n$}{
$est[i] = v_i + \frac{\watt - \floor*{\tilde{w}_i}}{\wdef - \floor*{\tilde{w}_i}}\sum\limits_{j \ne i} v_j$\;
}
$minI = \argmin(est)$\;
$\watt \minuseq \floor*{\tilde{w}_{minI}}$ , $\wdef \minuseq \floor*{\tilde{w}_{minI}}$\;
$\bm{v}.remove(minI)$ , $n \minuseq 1$\;
$bbEstRound(n, \bm{v}, \wdef, \watt)$\;

}
\caption{BBup: An efficient heuristic algorithm for discrete fractional setting}
\label{alg:bbup}
\end{algorithm}

Our next approach GD-f2c shown in Algorithm \ref{alg:gd-f2c} leverages Lemma \ref{lem:flcl}. The first condition therein implies that no weight $w_i$ is smaller than $\floor*{\tilde{w}_i}$. Hence, we initialize all $w_i$ to the said floor values (Line 2) and denote $s$ as the surplus weight to be distributed (Line 3). We then find the best item to absorb one unit of weight (Line 5-6) by greedily considering the global objective value computed using Dantzig's algorithm (Line 7-8). Once such an item is found and its weight incremented (Line 10-11), we repeat until all surplus is distributed.

\begin{algorithm}[h]
\DontPrintSemicolon
\SetKwProg{Def}{}{:}{}
$\bm{\tilde{w}} =$ VP weights\;
$\bm{w} = \left(\floor*{\tilde{w}_1}, \floor*{\tilde{w}_2}, \ldots, \floor*{\tilde{w}_n}\right)$\;
$s = \wdef - \sum_i w_i$\;
\For{$j = 1, 2, \ldots, s$}{
\For{$i = 1, 2, \ldots, n$}{
$\bm{w'} = (w_1, w_2, \ldots, w_i +1, \ldots wn)$\;
Sort $\bm{w'}$ in non-ascending BB\;
$est[i] = U_{\bm{w'}}(\watt)$\;
}
$minI = \argmin(est)$\;
$w_{minI} \pluseq 1$
}

\caption{GD-f2c: An efficient heuristic algorithm for discrete fractional setting}
\label{alg:gd-f2c}
\end{algorithm}

While the previous approach \textbf{g}reedily \textbf{d}istributes the surplus taking the weights from \emph{\textbf{f}loor values to \textbf{c}eiling} (hence the name GD-f2c), we analogously consider GD-c2f that initializes the weights to ceiling values and gradually deflates towards flow values, reflecting the second condition of Lemma \ref{lem:flcl}. Since only one of the two is guaranteed, we let GD+ to be the algorithm that considers the better of the two. By maintaining items in sorted order of BB, we are able to compute the Dantzig's algorithm in $O(n)$, and as the inner and outer loops run $O(n)$ and $O(s)$ times resp., the runtime complexity of these is $O(n^2 s)$. Note that while $s$ depends on item values, it is at most $O(n)$.

We now consider the 0-1 setting, i.e., the all-or-nothing scenario.

\section{0-1 Formulation}

Similar to the fractional setting, we have the following bi-level OP for the leader (0-1 model):
\begin{align*}
\min_{\sum w_i \le \wdef} \ \ \max_{x} \ & \sum_i v_ix_i \\
\text{s.t.} \ & \sum_i w_ix_i \le \watt \\
& x_i \in \{0,1\}
\end{align*}

In the fractional model, the last constraint was $x_i \in [0,1]$.
Recall that for the fractional knapsack formulation, Proposition 1 showed that the value-proportional weights are optimal.
We can show that this is not necessarily optimal in the 0/1 formulation as follows.
\subsection*{Counter-example}
Suppose we have two items with values 50 and 100. Suppose $\wdef = 15$ and $\watt = 12$. Setting the weights to 5 and 10 resp. leads to leader utility 100, but the optimal is 50 achieved by setting the weights, say, 2 and 13.
\par
Another one where the optimal solution does not set one cost entirely higher than $\watt$: Suppose we have 3 items with utility 10 and another 3 with 3. Suppose $\wdef = 390$ and $\watt = 235$. Setting the weights to 100 and 30 resp. for the two item types, yields a utility $23$ (as $100 + 100 + 30 \le 235$). But the optimal is $19$ achieved by setting the costs 120 and 10 resp. for the two item types.

\subsubsection*{Lower bound on objective} Notice that no matter how weights are set, the sum of the smallest $k$ weights will always fit in the knapsack, where $k = \floor*{\frac{\watt}{\wdef}n}$, leading to an objective value of $k$ smallest values summed. This bound is tight for several cases; in particular, we can outline the following propositions:

\begin{proposition}
    The lower bound above is tight when 
    \begin{itemize}
        \item $\frac{\watt}{\wdef} < \frac{1}{n-1}$, obtained by setting $w_i > \watt$ for all $i \ge 2$.
        \item $\ge 1 - \frac{1}{n-1}$, obtained by setting $w_n > \watt$. 
    \end{itemize}
    It follows that it is always tight for $n=2$.
\end{proposition}

\begin{table*}[t!]
\begin{tabular}{|c|cccc|ccc|}
\hline
n=2 & \multicolumn{4}{c|}{\begin{tabular}[c]{@{}c@{}}{[}0, 1/2)\\ 0\end{tabular}}                                                                                                                                                                                                                                                                                                           & \multicolumn{3}{c|}{\begin{tabular}[c]{@{}c@{}}{[}1/2, 1)\\ $v_1$\end{tabular}}                                                                                                                            \\ \hline
n=3 & \multicolumn{1}{c|}{\begin{tabular}[c]{@{}c@{}}{[}0, 1/3)\\ 0\end{tabular}} & \multicolumn{3}{c|}{\begin{tabular}[c]{@{}c@{}}{[}1/3, 1/2)\\ $v_1$\end{tabular}}                                                                                                                                                                                                                       & \multicolumn{1}{c|}{\begin{tabular}[c]{@{}c@{}}{[}1/2, 2/3)\\ $\min \{v_3, v_1+v_2\}$\end{tabular}} & \multicolumn{2}{c|}{\begin{tabular}[c]{@{}c@{}}{[}2/3, 1)\\ $v_1 + v_2$\end{tabular}}                \\ \hline
n=4 & \multicolumn{1}{c|}{\begin{tabular}[c]{@{}c@{}}{[}0, 1/4)\\ 0\end{tabular}} & \multicolumn{1}{c|}{\begin{tabular}[c]{@{}c@{}}{[}1/4, 1/3)\\ $v_1$\end{tabular}}     & \multicolumn{1}{c|}{\begin{tabular}[c]{@{}l@{}}{[}1/3, 2/5)\\ $\min \{v_3, v_1+v_2\}$\end{tabular}}          & \begin{tabular}[c]{@{}l@{}}{[}2/5, 1/2)\\ $\min \{v_4, v_1+v_2\}$\end{tabular}             & \multicolumn{1}{c|}{$\cdots$}                                                                               & \multicolumn{2}{c|}{\begin{tabular}[c]{@{}c@{}}{[}3/4, 1)\\ $v_1 + v_2 + v_3$\end{tabular}}                         \\ \hline
n   & \multicolumn{1}{c|}{\begin{tabular}[c]{@{}c@{}}{[}0, 1/n)\\ 0\end{tabular}} & \multicolumn{1}{c|}{\begin{tabular}[c]{@{}c@{}}{[}1/n, 1/(n-1))\\ $v_1$\end{tabular}} & \multicolumn{1}{l|}{\begin{tabular}[c]{@{}l@{}}{[}1/(n-1), 2/(2n-3))\\ $\min \{v_3, v_1+v_2\}$\end{tabular}} & \begin{tabular}[c]{@{}l@{}}{[}2/(2n-3), 1/(n-2))\\ $\min \{v_4, v_1+v_2\}$\end{tabular} & \multicolumn{1}{c|}{$\cdots$}                                                                               & \multicolumn{2}{c|}{\begin{tabular}[c]{@{}c@{}}{[}1-1/n, 1)\\ $v_1 + \ldots + v_{n-1}$\end{tabular}} \\ \hline
\end{tabular}
\caption{Table shows closed form expression for optimal objective for various interval ranges of $\frac{\watt}{\wdef}$ (top line in each cell) for various values of $n$ (specified in column 1).}
\end{table*}

Table 1 shows the closed form solutions for some special cases of the budget ratio. The values for $n \le 4$ are obtained with case-wise analysis, while they are extended to arbitrary $n$ for small budget values using an inductive argument. It is clear from column 1 that the sub-optimility of the VP weights is not bounded: if $\frac{v_1}{\bar{V}} \le \frac{\watt}{\wdef} < \frac{1}{n-1}$, the optimal is 0, whereas the VP weights yield non-zero as the follower can fit at least $v_1$ in the knapsack.

In addition to the lower bound above, we can also put an upper bound as the sum of the largest $k$ weights, since assigning equal weights to all items allows to fit at most $k$ items (with the follower picking the largest $k$), so the optimal solution is no worse.

\begin{proposition}
    There exists an optimal solution which satisfies $\forall i, j$:
    \begin{align*}
        v_i < v_j \Rightarrow w_i < w_j
        \quad\&\quad & v_i = v_j \Rightarrow w_i = w_j.
    \end{align*}
\end{proposition}

This result can be particularly beneficial to accelerate any (Bilevel) LP solvers by augmenting constraints as above.

\subsection*{Computational complexity for discrete allocation}

We now discuss the hardness of the problem as well as its containment in a complexity class.

We recall that the complexity class $\Sigma_2^P$ contains all decision problems that can be written in the form $\exists x \forall y P (x, y)$; that is, as a logical formula starting with an existential quantifier followed by a universal quantifier followed by a Boolean predicate $P(x, y)$ that can be evaluated in polynomial time; see, for instance, Chapter 17 in Papadimitriou’s book [23].
Just like the decision versions of bilevel problems DeRi [5], MACH, DNeg [6], our problem asks whether there exists a way of fixing the variables controlled by the leader, such that all possible settings of the variables controlled by the follower yield a good objective value for the leader.
Since this question is exactly of the form $\exists x \forall y P (x, y)$,
we conclude that it is contained in $\Sigma_2^P$. 

\begin{theorem}
    The problem is NP-hard.
\end{theorem}

\begin{proof}
    We reduce from the Number partition problem (NPP) which is known to be NP-hard. Suppose we are given an NPP instance ($v_1, v_2, \ldots, v_n$). We construct our knapsack game instance with $n$ items having these values, $\wdef = \sum_i v_i$ and $\watt = \wdef/2$. Then, we claim that the optimal partition has a discrepancy of $k$ IFF the knapsack game has the optimal objective $\watt-k/2$.

    To prove this, suppose the optimal partition is $(S, S')$ with $v(S) = v(S') - k$. Note that irrespective of the weights set by the leader, S and S' cannot both be unaffordable since their weights must sum to $\wdef$ and the the follower has a budget of $\watt = \wdef/2$. Thus, the optimal weight-assignment cannot achieve better than $\watt-k/2$.

    On the other hand, the value-proportional weight assignment achieves $\watt-k/2$ since S must the largest affordable subset (as a larger affordable subset will yield a smaller discrepancy for NPP). Thus, the optimal weight-assignment cannot achieve worse than $\watt-k/2$.

    Thus, the optimal in this case is precisely $\watt-k/2$.
\end{proof}
\subsubsection*{Minimmizing against local maxima}
We leverage the analysis from proposition 3 to infer that there is a k-sized knapsack solution that is \emph{nearly} optimal for the follower. Hence, we consider a locally optimal k-sized solution for the follower and minimize against that via the following MILP:

\begin{align}
\min_{\sum w_i \le \wdef, V} \ & \ V \\
\text{s.t.} \ & \sum_i w_ix_i \le \watt \\
& \sum_i v_ix_i \le V \\
& y_{jkj} = 0, y_{jkk} = 1 &\ \ \forall j \ne k \\
& y_{jki} = x_i \ \ \forall i \ne j,k &\ \ \forall j \ne k \\
& \sum_i w_iy_{jki} > \watt - M_1a_{jk} &\ \ \forall j \ne k\\
& \sum_i v_iy_{jki} \le V + M_2b_{jk} &\ \ \forall j \ne k\\
& a_{jk} + b_{jk} \le 1 &\ \ \forall j \ne k\\
& \sum_i x_i = k \\
& x_i, y_{jki}, a_{jk}, b_{jk} \in \{0,1\} & \ \ \forall i,j,k
\end{align}

Here, $x$ denotes the locally optimal solution we are looking for, characterized by constraints 2,3. Each $y_{jk}$ is a \emph{neighbor} of $x$ where item $j$ is (potentially) swapped with item $k$ as described by constraints 4,5. Constraints 6,7 ensure that if $y_{jk}$ fits in capacity, then $a_{jk} = 1$ and if $y_{jk}$ yields a higher value for the follower, then $b_{jk} = 1$. For $x$ to be locally optimal, constraint 8 ensures that there are no neighbors that are affordable as well as yielding more value.

The strict inequality in 6 can be handled by introducing a small constant $\epsilon$. The bilinear terms in 2,6 have a product of a binary and a continuous variable, which are known to be linearizable.

\section{Numerical results}

We now show numerical results evaluating our heuristics from section 4. We generate instances by sampling no. of items in the range $[10, 20]$ item values as partitions of total value $V$. We set the leader budget $\wdef = 100$ and to obtain fractional VP weights, we set $V = 100 \wdef$. We vary the follower budget by varying the ratio $\frac{\watt}{\wdef}$ in $\{0.1, 0.2, \ldots, 0.9\}$, and we choose 100 instances for each as described above. We compare our heuristic against the baseline of heuristic \emph{RR} that \textbf{r}ounds the VP weights \textbf{r}andomly.

\begin{figure*}[t]
\begin{subfigure}{.33\textwidth}
  \includegraphics[width=\linewidth]{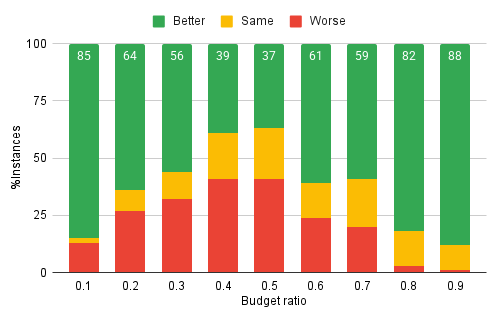}
  \caption{BBup}
  \label{fig:sub1}
\end{subfigure}
\begin{subfigure}{.33\textwidth}
  \includegraphics[width=\linewidth]{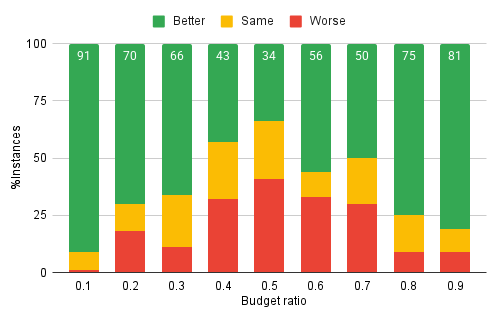}
  \caption{BBdown}
  \label{fig:sub2}
\end{subfigure}
\begin{subfigure}{.33\textwidth}
  \includegraphics[width=\linewidth]{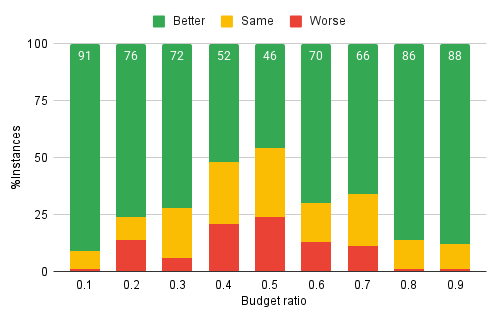}
  \caption{BB+}
  \label{fig:sub3}
\end{subfigure}

\medskip
\begin{subfigure}{.33\textwidth}
  \includegraphics[width=\linewidth]{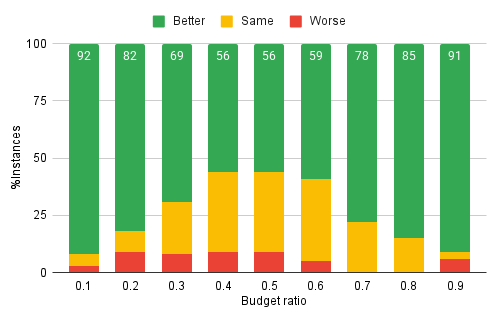}
  \caption{GD-f2c}
  \label{fig:sub4}
\end{subfigure}
\begin{subfigure}{.33\textwidth}
  \includegraphics[width=\linewidth]{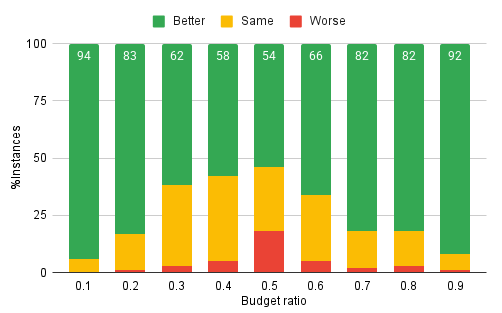}
  \caption{GD-c2f}
  \label{fig:sub5}
\end{subfigure}
\begin{subfigure}{.33\textwidth}
  \includegraphics[width=\linewidth]{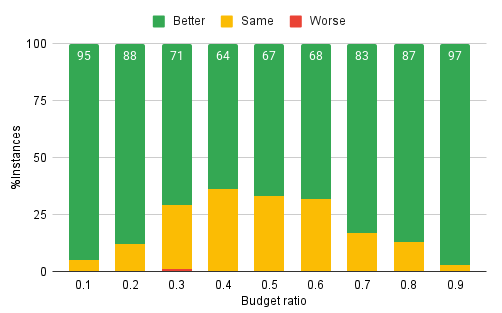}
  \caption{GD+}
  \label{fig:sub6}
\end{subfigure}
\caption{Comparison of our heuristics against the baseline of RR. Each plot shows the \%instances where our heuristic performs better than RR (in green), same (in yellow) and worse (in red) as budget ratio varies.}
\label{fig:numRes}
\end{figure*}

Our results are summarized in Fig. \ref{fig:numRes}. Each plot shows the no. of instances showing i) improvement due to our heuristic over RR (in green), ii) same output as RR (in yellow) and iii) worse output than RR (in red). We analyze the performance for varying values of the ratio $\frac{\watt}{\wdef}$ on x-axis.

Fig. \ref{fig:sub1}, \ref{fig:sub2}, \ref{fig:sub3} show the performance of BBup, BBdown and BB+ resp. against RR. We see that BBup, BBdown work effectively for budget ratio not close to 0.5, but not quite as well when it is close to 0.5. This is likely due to the fact that the linear estimation for the piecewise linear function used in our approach will naturally be more accurate for budget ratio closer to 0 or closer to 1. BB+ proves to be an effective remedy for this issue as BBup, BBdown are seen to have complementary efficacies. At its worst (for budget ratio 0.5) BB+ is worse than RR on 24\% instances, but improves output on 46\% instances. At its best (for budget ratio 0.1 or 0.9), these numbers resp. improve all the way to 1\% and \~90\%.

While GD-f2c, GD-c2f, and GD+ come with a slightly higher runtime complexity, they show significantly better output across the board. The former two individually out-perform BB+, and GD+ even more so with the bagging benefit. GD+ is worse than RR in a mere 1 among all the 1000 instances, while showing strict improvement in about 65\% instances at worst (for budget ratio around 0.5) and 97\% at best (for budget ratio 0.1 or 0.9)

\section{Conclusions}
This work explores competitive resource allocation in a sequential setting, diverging from the traditional simultaneous Colonel Blotto framework. We focus on a sequential decision-making scenario restricting to deterministic pure strategies that offers computational simplicity while maintaining strategic efficacy. Additionally, we expand the payoff structure to incorporate fractional allocations and payoffs, enabling more nuanced and proportional outcomes compared to the conventional binary framework.
We do so via a novel adversarial knapsack formulation, framing the problem as a bilevel optimization problem that integrates both the leader's and the follower's objectives. For the fractional allocation setting we show efficient heuristics where BB+ has the advantage of a $O(n^2)$ runtime complexity but manages relatively modest improvement in the worst case. The heuristic GD+ shows significant improvement even in the worst case for a slightly higher runtime complexity of $O(n^3)$ We also provide an analysis of the classical 0-1 knapsack case, showing to be NP-hard, closing a vital persisting gap in the literature while also providing an MILP heuristic that minimizes against local maxima.

\bibliographystyle{plain} 
\bibliography{references}

\end{document}